\documentclass[11pt, a4paper]{article}
\pdfoutput = 1
\usepackage[height=22cm,width=16cm, centering]{geometry}

\usepackage{setspace}

\usepackage[sort&compress, numbers, merge]{natbib}

\usepackage{amsmath, amssymb, mathrsfs,  comment}
\usepackage{ifpdf, xcolor, subcaption}
\usepackage[utf8]{inputenc}

\ifpdf        
  \usepackage{graphicx}     
  \usepackage[bookmarksopen,colorlinks=true, linkcolor=link_col,
  citecolor=cite_col, urlcolor=url_col,linktocpage=false]{hyperref}
\else     
  \usepackage[dvipdfmx]{graphicx}     
  \usepackage[dvipdfmx,bookmarksopen,colorlinks=true,linkcolor=link_col,
  citecolor=cite_col ,urlcolor=url_col,linktocpage=false]{hyperref}
\fi

\usepackage{multicol}
\definecolor{link_col}{rgb}{0.0, 0, 0.8}
\definecolor{cite_col}{rgb}{0.6, 0, 0.3}
\definecolor{url_col}{rgb}{0.6, 0, 0.3}

\def \ns{n_\text{s}}
\def \alphas{\alpha_\text{s}}
\def \betas{\beta_\text{s}}

\begin{document}

\begin{titlepage}
\begin{center}
\leavevmode \\

{\small 
\hfill KEK-TH-2034\\
\hfill KEK-Cosmo-221\\
}

\noindent
\vskip 1.5 cm
{\Large Primordial Black Hole Dark Matter and LIGO/Virgo Merger Rate \\ \vskip 0.5mm from Inflation with Running Spectral Indices:}\\ \vskip 0.3mm {\large Formation in the Matter- and/or Radiation-Dominated Universe}

\vskip 0.8 cm

{\Large  
 Kazunori Kohri$^{1,2,3}$ and Takahiro Terada$^1$}

\vskip 1. cm

{\textit {\scriptsize
$^1$Theory Center, IPNS, KEK, 1-1 Oho, Tsukuba, Ibaraki 305-0801, Japan\\
$^2$The Graduate University for Advanced Studies (SOKENDAI), 1-1, Oho, Tsukuba, Ibaraki 305-0801, Japan\\
$^3$Rudolf Peierls Centre for Theoretical Physics, The University of Oxford, 1 Keble Road, Oxford OX1 3NP, UK
}}

\vskip 1.6 cm

\begin{abstract}
{\normalsize 
We study possibilities to explain the whole dark matter abundance by primordial black holes (PBHs) or to  explain the merger rate of binary black holes  estimated from the gravitational wave detections by LIGO/Virgo.  
We assume that the PBHs are originated in a radiation- or matter-dominated era from large primordial curvature perturbation  generated by inflation.  We take a simple model-independent approach considering inflation with large running spectral indices which are parametrized by $n_\text{s}, \alpha_\text{s}$, and $\beta_\text{s}$ consistent with the observational bounds.
  The merger rate is fitted by PBHs with masses of $\mathcal{O}(10)$ $M_{\odot}$ produced in the radiation-dominated era. Then the running of running  should be $\beta_\text{s} \sim 0.025$, which can be tested by future observation.  On the other hand, the whole abundance of dark matter is consistent with PBHs with masses of  asteroids ($\mathcal{O}(10^{-17})~M_{\odot}$) produced in an early matter-dominated era if a set of running parameters are properly realized. 
}
\end{abstract}

\end{center}

\end{titlepage}

\section{Introduction}

After the first detection of gravitational wave which is directly
emitted from a merger of a binary black hole (BH) by LIGO/Virgo
collaboration~\cite{Abbott:2016blz}, interests in primordial black
holes (PBHs)~\cite{Hawking:1971ei, Carr:1974nx} have been
revived~\cite{Bird:2016dcv, Clesse:2016vqa, Sasaki:2016jop,
  Eroshenko:2016hmn, Carr:2016drx, Georg:2017mqk, Sasaki:2018dmp}.
The masses of BHs to fit the merger rate distribute at around $\sim
30 M_{\odot}$, where $M_\odot$ denotes the solar mass ($ =
2.0\times 10^{33}\, \text{g}$). The masses in this range are
higher than those of typical binary BHs (BBHs) formed in astrophysical
scenarios at the final stage of stellar evolution
of main sequence stars with solar metallicity
$Z_\odot$~\cite{Belczynski:2009xy, Spera:2015vkd,
TheLIGOScientific:2016htt}.
For somewhat lower metallicity $\mathcal{O}(0.1 \, \text{--} \, 0.01) Z_\odot$,
however, such heavy BBHs can be produced~\cite{2041-8205-715-2-L138, Belczynski:2016obo, Belczynski:2016jno}.
It is also possible that BBHs with masses of $\sim$ 30 $M_{\odot}$ can
be formed after deaths of Population~III (Pop.~III) stars with their
much lower metallicities~\cite{Kinugawa:2014zha}, but there still exist
large uncertainties in theoretical predictions of the merger rate, which  
conservatively amount to a factor of ${\cal O}(10^2)$ due to unknown
astrophysical parameters~\cite{Kinugawa:2015nla,
Belczynski:2016obo}.

On the other hand, PBHs can be produced in the early universe through
collapses of enhanced curvature perturbations~\footnote{For other
  mechanisms to produce PBHs, there may exist collapses of cosmic
  strings, critical collapses, collapses during phase transitions, and so
  on (see Ref.\cite{Carr:2009jm,Carr:2018rid} and references therein).} much before any compact stellar objects are formed. In this
case, the abundance of PBHs are calculated from the curvature
perturbation with less ambiguities~\cite{Carr:1975qj,Harada:2017fjm}
once a cosmological history is fixed. Then we can potentially explain
the BH merger rate inferred from the first LIGO/Virgo event, $2$--$53$ Gpc${}^{-3}$ yr${}^{-1}$~\cite{Abbott:2016nhf}, although only
formation rates to produce the BBHs have some uncertainties within a
couple of orders of magnitude~\cite{Bird:2016dcv, Clesse:2016vqa,
  Sasaki:2016jop, Ali-Haimoud:2017rtz}.  Because so far five BH merger
events have been detected by LIGO/Virgo~\cite{Abbott:2016blz,
  Abbott:2016nmj, Abbott:2017vtc, Abbott:2017oio, Abbott:2017gyy},
this motivation becomes firmer.  Moreover, PBHs serve as a candidate
of ``non-baryonic'' cold dark matter (CDM).  Thus far a lot of people have discussed this possibility for a variety of mass ranges, \textit{e.g.}, see Refs.~\cite{Chapline:1975,Carr:1975qj,GarciaBellido:1996qt,Jedamzik:1999am,Frampton:2010sw,Kawasaki:2012wr,Kohri:2012yw,Frampton:2015xza, Axelrod:2016nkp,Carr:2016drx,Ezquiaga:2017fvi,Clesse:2017bsw} and references therein.  Also, there is a scenario in which DM is produced by Hawking radiation of PBHs~\cite{Allahverdi:2017sks, Lennon:2017tqq}.  PBHs are an interesting
dark matter (DM) candidate from the particle physics perspective
because we do not have to introduce new degrees of freedom beyond the
Standard Model to explain DM.

In this paper, we explain the BH merger rate or the entire DM
abundance by PBHs, but not both at the same time. We consider
inflation with ``running spectral indices'', 
\begin{align}
\alphas =& \frac{\partial \ns}{\partial \ln k}, &  \betas = \frac{\partial^2 \ns}{\partial (\ln k)^2},
\end{align}
where $k$ is the wave number, and $\ns$ is the spectral index of the primordial curvature perturbation.
These parameters give a quite simple phenomenological and
model-independent descriptions of the power spectrum of the curvature
perturbations. A large positive running, which is sometimes realized
by a large ``running of running'' $\betas$, can predict a large fluctuation at
small scales in which we are
interested~\cite{Green:1997sz,Covi:1998jp,Leach:2000ea,Kohri:2007qn,Bugaev:2008gw,Alabidi:2009bk,Drees:2011hb,Drees:2011yz,Alabidi:2012ex,Alabidi:2013wtp}.
In earlier works including Refs.~\cite{Clesse:2015wea, Cheng:2016qzb,
  Kawasaki:2016pql, Garcia-Bellido:2016dkw, Inomata:2016rbd,
  Inomata:2017okj, Garcia-Bellido:2017mdw, Ezquiaga:2017fvi,
  Kannike:2017bxn, Carr:2017edp, Ballesteros:2017fsr, Inomata:2017uaw,
  Inomata:2017bwi, Ando:2017veq, Hertzberg:2017dkh, Cheng:2018yyr}, the BH merger
rate for LIGO/Virgo events and/or the DM abundance in inflationary
scenarios have been discussed.  Here we do not explicitly introduce 
features like double inflation~\cite{Kannike:2017bxn, Inomata:2017bwi}
or additional fields like a curvaton~\cite{Carr:2017edp,
  Inomata:2017uaw, Ando:2017veq} to explain DM or LIGO/Virgo events.
It is remarkable and encouraging that such a simple inflationary power
spectrum can account for the LIGO/Virgo BH merger rate or the whole
DM.  The former (latter) is explained by the PBHs of intermediate
masses $M \sim \mathcal{O}(10)M_\odot$ (asteroid masses
$M \sim \mathcal{O}(10^{-17})M_\odot$).  Another feature
of this paper is that we consider both cases of BH formation in the
radiation-dominated (RD) era and in an early matter-dominated (MD) era.
The latter is less extensively studied in the literature, but it is
well-motivated in the inflationary cosmology since \textit{e.g.}~the
coherent oscillation phase of the inflaton  before
reheating behaves as a MD era.  For the probability of PBH formations
in a MD era, we take into account the effects of
anisotropies~\cite{Khlopov:1980mg,Polnarev:1982,Harada:2016mhb} and
angular momentum~\cite{Harada:2017fjm}, which suppress the PBH
formation compared to the case without these
effects~\cite{Khlopov:1980mg, Polnarev:1986bi}.  Differences of this
paper from Ref.~\cite{Carr:2017edp} include the facts that we also
consider the running of running parameter $\betas$, that we take into
account the effects of angular momentum on the PBH formation rate in
the MD era, and that we do not rely on a spectator field.
 
In the next section, we introduce parametrization of the power
spectrum of the curvature perturbations with the parameters for the
running spectral indices.  In Sec.~\ref{sec:formation}, the
probability of the PBH formation is reviewed in a RD or a MD era, and
its relation to the current abundance of PBHs is introduced.
Observational constraints on PBH abundance and curvature
perturbations are summarized in Sec.~\ref{sec:constraints}.  The main
part of the paper is written in Sec.~\ref{sec:results} where we scan
running parameters and find allowed regions to fit either the
LIGO/Virgo events or all the DM abundance.  The conclusions are given
in Sec.~\ref{sec:conclusion}.

\section{Inflation with running parameters}\label{sec:running}
We assume that the origin of the primordial curvature perturbations
needed for the PBH formation is the same as that produced by the inflaton
perturbations.  We take a phenomenological approach which is
independent of details of inflation models.  Then, we simply
parametrize the primordial curvature perturbations as
\begin{align}
P_\zeta (k) = A_\text{s} \left( \frac{k}{k_*} \right)^{n_\text{s} -1 + \frac{\alpha_\text{s}}{2} \ln \left( \frac{k}{k_*} \right)+ \frac{\beta_\text{s}}{6} \left(\ln \left( \frac{k}{k_*} \right)\right)^2 }, \label{eq:pzeta0}
\end{align}
where $A_\text{s}=( 2.207\pm 0.074) \times 10^{-9}$ (Planck 2015 TT,TE,EE+lowP (68\% CL)) is
the overall normalization~\cite{Ade:2015xua}, $k_*=0.05 \text{Mpc}^{-1}$ is the pivot
scale, $n_\text{s}$ is the spectral index, and $\alpha_\text{s}$ is
its running, and $\beta_\text{s}$ is its running of running.  The
parametrization considering higher-order corrections up to
$\beta_\text{s}$ is used by the Planck
collaboration~\cite{Ade:2015lrj}. Thus, we can compare those
parameters with observations of cosmic microwave background (CMB).

Because the scales of the wave number $k$ which are typical for CMB
observations and for productions of the PBH are different from
each other, this parametrization may have limitations partly on the
use of comparisons.  For example, if it is the case of
$\ln(k/k_*) \gg 1$, higher-order runnings  could become
more important. Their magnitudes depend on details of an inflaton
potential or possible additional light fields like a curvaton.  After
emphasizing this fact, we take simplicity instead of generality.  Our
parameters can be directly compared with CMB, and we do not introduce
additional ingredients explicitly such as double inflation or
spectator fields (curvatons).  For definiteness, we assume the
curvature perturbations are adiabatic and Gaussian.  We also assume
for simplicity that inflation ends instantaneously.  The last
assumption would be an aggressive one for the purpose only to avoid
constraints, but would be conservative when one tries to explain the
dark matter abundance or the merger rate.  
We will briefly discuss to what extent we can relax this assumption.
When the e-folding number during the inflation is small, another second inflation should follow the (first) inflation which produced perturbations.\footnote{
Typically, 50 to 60 e-foldings are required depending on the scale of inflation and the details of reheating.
For example, if the e-folding number of the first inflation with the running parameters is 30, a second inflation with 20 to 30 e-foldings is implicitly assumed.
}

From the current observations, the running parameters $\alphas$ and
$\betas$ are surely consistent with zero, but can also take finite
values.  A negative value of $\alphas$ is favored when $\betas$ is
turned off, but this tendency no longer remains when the latter is
turned on. The current constraints on these parameters due to Planck
2015 TT, TE, EE+lowP (68\% CL) are as follows~\cite{Ade:2015lrj},
\begin{align}
\ns =& 0.9586 \pm 0.0056 , \\
\alphas =& 0.009 \pm 0.010, \\
\betas =& 0.025 \pm 0.013.
\end{align}
We  adopt these bounds on the parameters.\footnote{
The Planck 2018 (TT, TE, EE$+$lowE$+$lensing) constraints~\cite{Akrami:2018odb}, which were released after the completion of this paper, are
\begin{align}
n_{\text{s}}= & 0.9625 \pm 0.0048 , \\
\alphas =& 0.002 \pm 0.010, \\
\betas =& 0.010 \pm 0.013.
\end{align}
The values of the parameters used in this paper are consistent with these constraints except for a small portion of Fig.~\ref{fig:f_contour}, which in any case cannot produce a substantial amount of PBHs.  In particular, the benchmark points for LIGO/Virgo (Fig.~\ref{fig:fLIGO}) and for dark matter (Fig.~\ref{fig:f=1}) are still well within the 2$\sigma$ bound.
}

As will be shown later, we definitely need a finite positive value of
$\betas$ to fit the LIGO/Virgo merger rate by the PBHs produced at
around $k = k_{\rm LIGO} \sim 10^6$~Mpc$^{-1}$. The curvature
perturbation is required to be
$P_{\zeta}(k_{\rm LIGO} ) \sim 3\times 10^{-2}$ at $k=k_{\rm LIGO}$.
By putting $P_{\zeta}(k_{\rm *} ) = A_s = 2.2 \times 10^{-9}$ and
$N(k_{\rm LIGO}) \equiv \ln\left(k_{\rm LIGO} /k_{\rm *} \right) \sim 17$ into
Eq.~(\ref{eq:pzeta0}) with $n_s -1 \sim -0.04$, we find a relation of
the condition to produce the PBHs to be
  \begin{eqnarray}
    \label{eq:alpha+beta}
-0.04 \left(\frac{n_s-1}{-0.04} \right) \left( \frac{N(k_{\rm LIGO})}{17}\right)
+ 0.1 \left( \frac{\alpha_s}{0.01}\right) 
     \left( \frac{N(k_{\rm LIGO})}{17}\right)^2
+ 0.1 \left( \frac{\beta_s}{0.002}\right) 
     \left( \frac{N(k_{\rm LIGO})}{17}\right)^3
\sim 1.
\end{eqnarray}
For the observational constraint on the running,
$\left| \alphas \right| \lesssim 0.01$, we approximately need a positive running of running  with the order of $\betas \sim 0.02$.
On the other hand, PBHs for dark matter require smaller values for the running parameters, as we will see in the subsequent sections.

The running parameters are constrained also by supernovae lensing~\cite{Ben-Dayan:2013eza, Ben-Dayan:2015zha}.
It involves nonlinear evolution and astrophysical uncertainties, so we do not adopt such bounds here.
Nevertheless, it is easy to compare our results with such constraints.

\section{PBH formation probability}\label{sec:formation}
The fraction of PBHs in the energy density at the time of PBH formation is conventionally denoted by $\beta$, which should not be confused with the running of running $\betas$.
The fraction $\beta$ can also be interpreted as the probability for a given Hubble patch to become a PBH.
It depends on the equation of state of the universe.
\subsection{PBH formation in a RD era}
To consider the PBH formation, it is appropriate to smooth out
subhorizon modes because it is largely determined by the horizon mass
but not by tiny structures inside the horizon
approximately. (For more precise computations however, we need
  details of the profile for the density
  perturbation~\cite{Nakama:2013ica,Nakama:2014xwa}.) 
We define a coarse grained density perturbation $\sigma$~\cite{Young:2014ana},
\begin{align}
\sigma^2(k) = & \int_{-\infty}^\infty \text{d} \ln q \, w^2\left(\frac{q}{k} \right) \frac{4(1+w_{\text{eos}})^2}{(5+3w_{\text{eos}})^2} \left(\frac{q}{k}\right)^4 P_\zeta (q), \label{sigma(k)}
\end{align}
where $w(k)=\exp (-k^2/2)$ is a Gaussian window function, and $w_{\text{eos}}=P/\rho$ is the equation-of-state parameter ($w_{\text{eos}}=1/3$ in the RD era; $P$ and $\rho$ are the pressure and the energy density).
In the above expression, an weighted average is taken with respect to the wavenumber $q$ with a smaller weight for a large $q(\gtrsim k)$ according to the window function. 
The transfer function of the density perturbations has been neglected because it is not important for the Gaussian window function~\cite{Ando:2018qdb}.  
The $\sigma$, encoding the information of the power spectrum $P_\zeta$, measures the typical strength (standard deviation) of density perturbations.

In the Press-Schechter formalism~\cite{Press:1973iz}, a PBH forms just after the density perturbation $\delta$ larger than a critical value $\delta_c$ enters the Hubble horizon.  Assuming that the density perturbation $\delta$ obeys the Gaussian distribution with the variance $\sigma^2$,  this criterion means that the PBH formation probability is given by
\begin{align}
\beta (\sigma) = \int_{\delta_\text{c}}^{\infty} \frac{1}{\sqrt{2\pi}N\sigma} \exp \left(\frac{-\delta^2}{2\sigma^2}\right) \text{d}\delta \simeq \frac{1}{2} \text{Erfc}\left( \frac{\delta_{\text{c}}}{\sqrt{2}\sigma} \right) ,
\end{align}
where  $N=\int_{-1}^{\infty} \frac{1}{\sqrt{2\pi}\sigma} \exp{\frac{\delta}{2\sigma^2}} \text{d}\delta$ is the normalization factor.
The critical value $\delta_\text{c}$ is $1/3$ in a simple
analytic derivation~\cite{Carr:1975qj} and 0.42 - 0.56 in more
sophisticated approaches~\cite{Shibata:1999zs, Musco:2004ak,
  Polnarev:2006aa, Musco:2008hv, Musco:2012au, Nakama:2013ica, Harada:2015yda}.  We
take $\delta_c = 0.42$ as a reference value in this paper.  The lower end of the integral of $N$ is set to $-1$ since perturbations with $\delta < -1$ is not produced. 
However, $\sigma \ll 1$ in our relevant parameter space, which means the integral is dominated at $|\delta|\ll 1$.  Therefore, $N$ is approximated to  unity in the second equality.

To reduce the calculation cost, we approximate $\sigma$ to be
\begin{align}
\sigma^2 (k) = \frac{2(1+w_{\text{eos}})^2}{(5+3w_{\text{eos}})^2} P_\zeta (k),
\end{align}
which is exact when $P_\zeta(k)$ is scale invariant.  Even when the
running parameters are introduced, the ratio of the exact and the
approximated $\sigma^2$ is roughly within an order of magnitude.  This
is within the same magnitude as that of uncertainties coming from the
choice of the window function.

\subsection{PBH formation in a MD era}
In the MD era ($w_{\text{eos}}=0$), density fluctuations grow
in proportion to the scale factor once the scales of the corresponding
wavelengths enter the Hubble horizon.  In contrast to the RD case,
initially small fluctuations can become large and eventually collapse
to a PBH if the MD era is sufficiently long~\cite{Khlopov:1980mg,
  Polnarev:1986bi}.  This effect significantly enhances the formation probability of PBHs
compared to the case of the RD era.  However, effects of
anisotropies~\cite{Harada:2016mhb} and accumulating angular momentum
of the perturbed region~\cite{Harada:2017fjm} suppress the formation
probability.  The analytic expression is approximately given by the
following formula~\cite{Harada:2017fjm},
\begin{align}
\beta (\sigma) = \begin{cases}
1.894\times10^{-6} \times f_Q \mathcal{I}^6 \sigma^2 \exp \left(-0.1474 \frac{\mathcal{I}^{4/3}}{\sigma^{2/3}}\right) & (\sigma < 0.005) \\
0.05556 \sigma^5 & (\sigma \geq 0.005)
\end{cases}
\end{align}
where $\mathcal{I}$ is a dimensionless variable characterizing the
magnitude of the angular momentum of the system, $Q$ is a
dimensionless parameter measuring the initial quadrupole moment of the
mass, and $f_Q$ is the fraction of masses whose $Q$ is below a
critical value. The continuity of $\beta$ leads to $f_Q \simeq 0.57$
for $\mathcal{I}=1$.  In our calculation, we use more precise
numerical data to actually plot $\beta(\sigma)$ in Fig.~5 of
Ref.~\cite{Harada:2017fjm} instead of the above formula.

\subsection{Current PBH abundance}
We consider the present value (or the value just before the PBHs evaporate) of the fraction 
  $f_{\text{PBH}}$ of the energy density of PBHs to that  of cold dark
matter (CDM), which is defined as
\begin{align}
f_{\text{PBH}}=& \frac{\rho_{\text{PBH}}}{\rho_{\text{CDM}}}, \label{f_definition}
\end{align}
where $\rho_X$ ($X=$PBH, CDM) is the energy density of PBH or CDM.
We introduce the differential energy density and fraction by
\begin{align}
\rho_{\text{PBH}}(M)=& \frac{\text{d}\rho_{\text{PBH}}}{\text{d}\ln (M/M_\odot)},  &  f_{\text{PBH}}(M) =& \frac{\text{d} f_{\text{PBH}}}{\text{d} \ln (M/M_\odot) }, 
\end{align}
so that the total fraction is obtained as a logarithmic integral,
\begin{align}
f_{\text{PBH}}=&\int \text{d}\ln (M/M_\odot) \,  f_{\text{PBH}}(M). \label{f_integrate}
\end{align}

The fraction $f_{\text{PBH}}(M)$ can be calculated by using the PBH formation probability $\beta(\sigma)$ introduced in the previous subsections. Then we obtain
\begin{align}
f_{\text{PBH}}(M)=\left. \left( \frac{g_{*}(T)}{g_{*}(T_{\text{eq}})} \frac{g_{*,s}(T_{\text{eq}})}{g_{*,s}(T)} \frac{T}{T_{\text{eq}}} \gamma \beta(\sigma(k(M))) \right)\right |_{T= \text{Min}[T_M, T_{\text{R}} ]} \frac{\Omega_{\text{m}}}{\Omega_{\text{CDM}}}, \label{fandbeta}
\end{align}
where $g_{*} \, (g_{*,s})$ is the effective relativistic degrees of
freedom for energy (entropy) density, $T_{\text{eq}}$ is the
temperature at the (later) matter-radiation equality, $\Omega_{X}$
($X=$ m, CDM) is the energy density fraction of non-relativistic matter
or CDM, and we used
$\gamma \beta = \frac{\rho_{\text{PBH}}(M)}{\rho_{\text{total}}(T)}$
with $\gamma$ denoting the fraction of the horizon mass which actually
enters the PBH.  Explicitly, the definition of $\gamma$ is given by
the following relation
\begin{align}
M = \gamma \times \frac{4\pi}{3} H^{-3} \rho. \label{horizon_mass}
\end{align}
In a RD era, a simple analytic formula $\gamma = (1/\sqrt{3})^3$ is
known~\cite{Carr:1975qj}, while we set $\gamma=1$ in a MD era.  The expression inside the large parenthesis on the right-hand side of eq.~\eqref{fandbeta} is evaluated at the PBH formation,
which is $T=T_M$ in a RD era where $T_M$ denotes the temperature at
which the mode corresponding to the mass $M$ enters the Hubble
horizon, and $T=T_{\text{R}}$ (reheating temperature) dominates in the case of PBH formation
in a MD era.  The relation between mass $M$ and wavenumber $k$ is
obtained by Eq.~\eqref{horizon_mass} supplemented with $k=a H$ and the
Friedmann equation $3 M_{\text{P}}^2 H^2 =\rho $ where $M_{\text{P}}$
is the reduced Planck mass.  The wavenumber is further related to the
coarse-grained perturbation $\sigma(k)$ by Eq.~\eqref{sigma(k)}.

\subsection{Required amount of PBHs}
Before proceeding, we summarize the requirement for the PBH abundance.
The condition for the 100\% dark matter abundance is simple: $f_{\text{PBH}}=1$.
The condition to explain the merger rate requires some explanation.

First, the LIGO/Virgo collaboration estimated the merger rate of BBHs as $2$--$53 \, \text{Gpc}^{-3} \text{yr}^{-1}$ at the 90\% confidence level based on the event GW150914~\cite{Abbott:2016nhf}.  It was assumed that all the BHs have the same mass and spin with those of this event. 

To discuss the criterion for the merger rate, we introduce the PBH fraction $f_{\text{PBH}}^{\text{LIGO}}$
restricted to the LIGO/Virgo mass range to discuss the BH merger rate.
For definiteness, we define the range as $4 \leq M/M_\odot \leq 40$, then
\begin{align}
f_{\text{PBH}}^{\text{LIGO}}=\int_{\ln 4}^{\ln 40} \text{d}\ln (M/M_\odot) \, f_{\text{PBH}}(M). \label{f_LIGO}
\end{align}

There are largely two scenarios to produce binary systems from the PBHs.  In one scenario, a BBH ``forms'' in the RD era when the mass of the close pair of BHs become dominant compared to the energy of surrounding radiation. This scenario requires $10^{-3}\leq f_{\text{PBH}}^{\text{LIGO}}\leq 10^{-2}$ to explain the merger rate~\cite{Sasaki:2016jop}.  In the other scenario, a BBH forms when two PBHs encounter with a small enough impact parameter in the late-time Universe.  This mechanism is less efficient and requires $f_{\text{PBH}}\sim 1$~\cite{Bird:2016dcv}, which is in strong tension with various constraints for the $M \sim 30 M_{\odot}$ mass range.  Therefore, we adopt the former criterion, $10^{-3}\leq f_{\text{PBH}}^{\text{LIGO}}\leq 10^{-2}$.

Let us estimate in passing the number density of BBHs $n_{\text{BBH}}$ in the present Universe.
Using eq. (7) and the condition below eq. (3) in Ref.~\cite{Sasaki:2016jop}, the probability $R$ for a close pair of BHs to form a binary is evaluated as 
\begin{align}
R =& \frac{\int_0^{f_{\text{PBH}}^{\text{LIGO}}{}^{1/3}\bar{x}} dx \int_x^{\bar{x}} dy \frac{9}{\bar{x}^6} x^2 y^2}{\int_0^{\bar{x}} dx \int_x^{\bar{x}} dy \frac{9}{\bar{x}^6} x^2 y^2}
 = f_{\text{PBH}}^{\text{LIGO}} (2 - f_{\text{PBH}}^{\text{LIGO}}) \approx 2 f_{\text{PBH}}^{\text{LIGO}}  ,
\end{align}
where $x$ is the distance between the BHs in the binary at the time of matter-radiation equality,  $y$ is the distance of the binary to the nearest third BH, and $\bar{x}$ is the mean separation of BHs.  In the last equality, we approximated the formula assuming $f_{\text{PBH}}\ll 1$. 
Then $n_{\text{BBH}}$ is estimated as
\begin{align}
n_{\text{BBH}} \simeq & \frac{n_{\text{PBH}}R}{2}  = 3H^2 M_{\text{P}}^2 f_{\text{PBH}}^{\text{LIGO}}{}^2 \frac{\Omega_{\text{CDM}}}{M} \nonumber \\
& = 2\times 10^2 \, \text{Mpc}^{-3} \left( \frac{f_{\text{PBH}}^{\text{LIGO}}}{10^{-3}} \right)^2  \left( \frac{\Omega_{\text{CDM}}h^2}{0.12} \right)  \left( \frac{M}{30 M_{\odot}} \right)^{-1}.
\end{align}

\section{Observational constraints}\label{sec:constraints}

Observational constraints on the abundance of PBHs $f_{\text{PBH}}(M)$
and the curvature fluctuations $P_\zeta (k)$ are briefly reviewed in
this section.  PBHs lighter than
$M_{\text{evap}}=2.6 \times 10^{-19}M_{\odot}$ have been evaporated by
Hawking radiation~\cite{Hawking:1974sw}.  The emitted radiation
affects big-bang nucleosynthesis (BBN)~\cite{Carr:2009jm}, the cosmic
microwave background~\cite{Carr:2009jm, Stocker:2018avm} and the
galactic gamma-ray background~\cite{Carr:2009jm}, which in turn
severely constrain the PBH abundance.  PBHs slightly heavier than
$M_{\text{evap}}$ also emit energetic gamma ray, and they are
constrained by the extragalactic gamma-ray
background~\cite{Carr:2009jm}.

Heavier PBHs can be probed by gravitational lensing including
femtolensing of gamma-ray bursts~\cite{Barnacka:2012bm}, microlensing
constraints of Subaru/HSC~\cite{Niikura:2017zjd},
EROS-2~\cite{Tisserand:2006zx}, and MACHO~\cite{Allsman:2000kg}, and
caustic crossing~\cite{Oguri:2017ock}.  The constraint of Subaru/HSC
turned out to be invalid for $M\lesssim 10^{-10}M_{\odot}$ because the
geometric optics approximation is no longer valid~\cite{TakadaTalk,
  Nakamura:1997sw, Takahashi:2003ix} (see also
Ref.~\cite{Inomata:2017bwi}). Thus for the moment we cut the
constraint below $10^{-10}M_{\odot}$ by hand.  There are also
constraints from dynamical processes such as destruction of white
dwarfs by PBHs~\cite{Capela:2012jz, Graham:2015apa} and absorption of
neutron stars by PBHs~\cite{Capela:2013yf}.
Constraints for $10^{-5}M_{\odot}\lesssim M \lesssim 10^{-1}M_{\odot}$ could be potentially much improved, $f_{\text{PBH}} \lesssim 10^{-3} \sim 10^{-4}$, by gravitational waves from PBH-super massive BH binaries in future~\cite{Guo:2017njn}.

\begin{figure}[tbh]
 \centering
\includegraphics[width=0.5 \columnwidth]{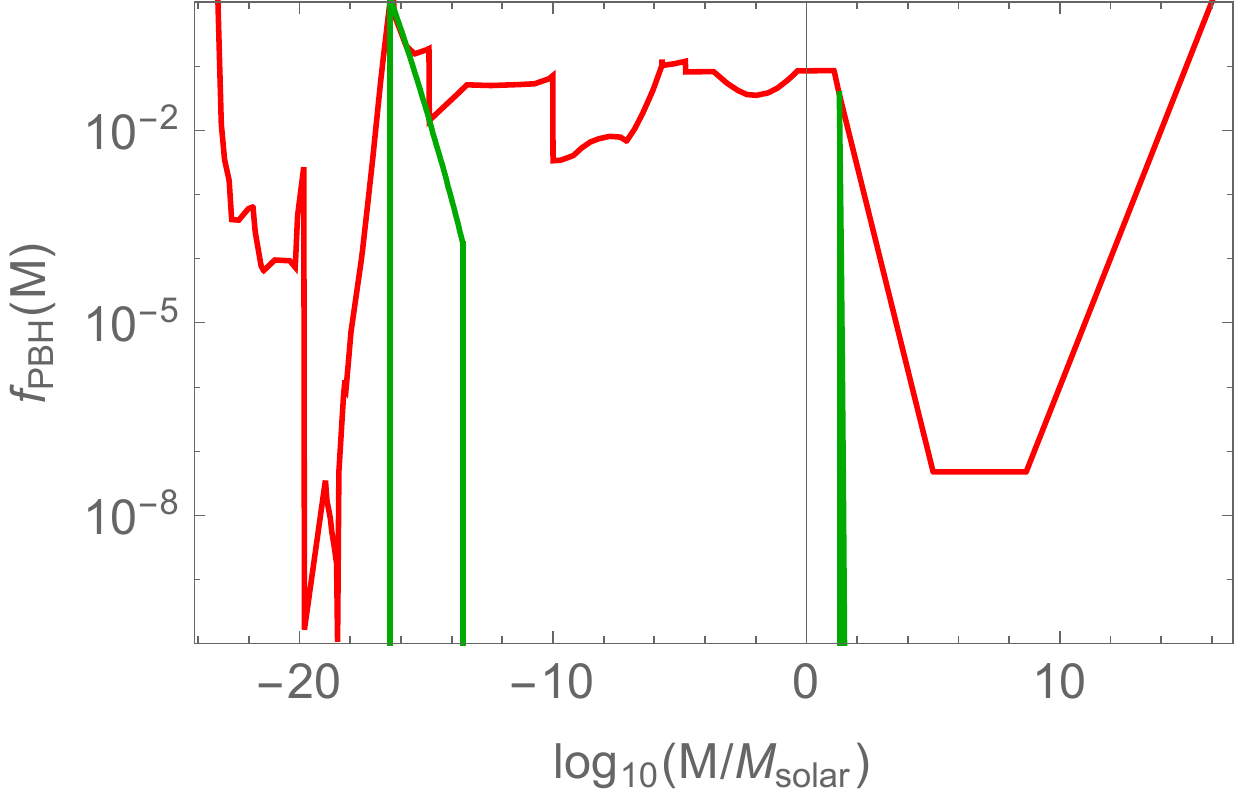}
  \caption{Constraints on the abundance of PBHs for the monochromatic mass function shown by the continuous red line.  The single-peaked green lines give two independent examples of the mass function.  The left one at around $10^{-16.5} - 10^{-13.5} M_{\odot}$ (from Fig.~\ref{fig:f=1}\subref{sfig:f=1})  explains the whole dark matter, and the right one  at around $\sim 10^{1}  M_{\odot}$  (from Fig.~\ref{fig:fLIGO}\subref{sfig:fLIGO}) explains the BH merger rate, respectively.}
 \label{fig:f(M)constraints}
  \end{figure}

For larger masses ($M\gtrsim 10^2 M_\odot$), accretions of baryonic
matter onto PBHs give bounds.  Because baryonic matter emits
high-energy photons during the accretion, reionization histories of
atoms and thermal histories of the Universe are significantly
modified. Then, CMB photons are affected by those high-energy
photons. From observations of fluctuations and polarization for the
CMB photons, the abundance of PBHs can be
constrained~\cite{Ali-Haimoud:2016mbv, Poulin:2017bwe}.  In
particular, non-spherical disk-accretions occur inevitably due to a
finite relative velocity between PBHs and baryonic matter. For the
non-spherical nature of the accretion disks, energy deposition would
be smaller, however, the reionization fraction becomes larger than the
cases of spherical accretions. Then, PBHs heavier than the solar mass
are severely constrained by observation~\cite{Poulin:2017bwe}.
Other bounds also come from radio and X-ray
observations~\cite{Gaggero:2016dpq, Inoue:2017csr}.  Gravitational
lensing of supernovae by PBHs put an independent constraint on the
abundance of PBHs~\cite{Zumalacarregui:2017qqd,
  Garcia-Bellido:2017imq}.  Furthermore, PBHs are constrained by
formations of large scale structure~\cite{Carr:2018rid}.  There are
yet more constraints in these high mass range including survival of
stars in ultra-faint dwarf galaxies (Segue
I~\cite{Koushiappas:2017chw} and Eridanus II~\cite{Brandt:2016aco}),
wide binaries~\cite{Monroy-Rodriguez:2014ula}, globular
clusters~\cite{Carr:2009jm}, and so on.  Not all of them are robust,
but the presence of a number of independent constraints with different
physical requirements indicate robustness when taken together.
(However, see also Ref.~\cite{Green:2017qoa} discussing astrophysical
uncertainties for constraints on multi-solar mass PBHs.)  Finally,
PBHs should not exceed the total dark matter abundance,
$f_{\text{PBH}}\leq
1$~\cite{Ade:2015xua}. 
These constraints on $f_{\text{PBH}}(M)$ are summarized in
Ref.~\cite{Carr:2009jm,Carr2018prep} and combined in
Fig.~\ref{fig:f(M)constraints}.

To produce PBHs, large curvature perturbations are required, especially in a RD era.
Such large perturbations are directly constrained by $\mu$-distortion of CMB~\cite{Chluba:2012we, Kohri:2014lza} and the so-called ``acoustic reheating'' at BBN~\cite{Jeong:2014gna,Nakama:2014vla, Inomata:2016uip}.
In addition, large scalar perturbations induce gravitational waves in the second order of perturbations~\cite{Ananda:2006af, Baumann:2007zm}.
They are constrained by observations~\cite{Saito:2008jc, Saito:2009jt,Alabidi:2012ex,Alabidi:2013wtp}, \textit{e.g.},  through pulsar timing array (PTA) experiments~\cite{Lentati:2015qwp, Shannon:2015ect, Arzoumanian:2018saf}.

Before closing this section, it is worth mentioning that these constraints are derived for the monochromatic mass function.
For discussions on the cases with an extended mass function, see Refs.~\cite{Carr:2016drx, Green:2016xgy, Kuhnel:2017pwq, Carr:2017jsz, Bellomo:2017zsr, Lehmann:2018ejc}.
We adopt the method in Ref.~\cite{Carr:2016drx} to derive an appropriate constraint for our extended mass function when necessary. 

\section{Parameter scan results}\label{sec:results}

We study the dependence of the fraction $f_{\text{PBH}}$ of PBHs in
dark matter on the parameters of the curvature perturbation with the
running spectral indices.  This is obtained after integrating
$f_{\text{PBH}}(M)$ over the mass as in Eqs.~\eqref{f_integrate} and \eqref{f_LIGO}.

More detailed procedure is as follows.  We consider five parameters
($\ns$, $\alphas$, $\betas$, $T_{\text{R}}$, $H_{\text{MD}}$) where
$H_{\text{MD}}$ is the Hubble value when the early MD era begins.  The
end of the MD era, on the other hand, is determined by the reheating
temperature $T_{\text{R}}$.  For each parameter set, the wavenumber
$k$ is scanned over 18 orders of magnitude from the CMB scale to the
scale corresponding to the left edge of the constraint in
Fig.~\ref{fig:f(M)constraints}.  For each $k$, the power spectrum
$P_\zeta (k)$ or the mass function $f_{\text{PBH}}(M)$ is compared to
the corresponding constraint.  When the wavenumber becomes larger than
a critical scale\footnote{ Roughly speaking, the critical scale
  corresponds to the time of reheating when the early MD era ends and
  the RD era begins.  Since the modes entering the horizon at the last
  stage of the MD era do not have enough time to become nonlinear, the
  constraint becomes weak.  For such modes, we adopt the constraint
  for the RD era.  }, the calculation of the PBH abundance assuming RD
is taken over by that assuming MD.  When the $P_\zeta(k)$ or
$f_{\text{PBH}}(M)$ touches the constraint, we stop the scan of $k$
(or equivalently $M$).  This is because our power spectrum has a
rising shape due to positive running parameters, 
 and thus it will be excluded soon when
$k$ is further increased even when the constraints are for the
monochromatic mass function.  
 Then we integrate the mass function $f_{\text{PBH}}(M)$ in the relevant region.  Because
of the positive running in most of our parameter space, the
integration is dominated by the small scale (large $k$, small $M$), and thus
it is insensitive to the upper end of the integral. The results are
presented below.

\begin{figure}[tbh]
 \centering
\includegraphics[width=0.5 \columnwidth]{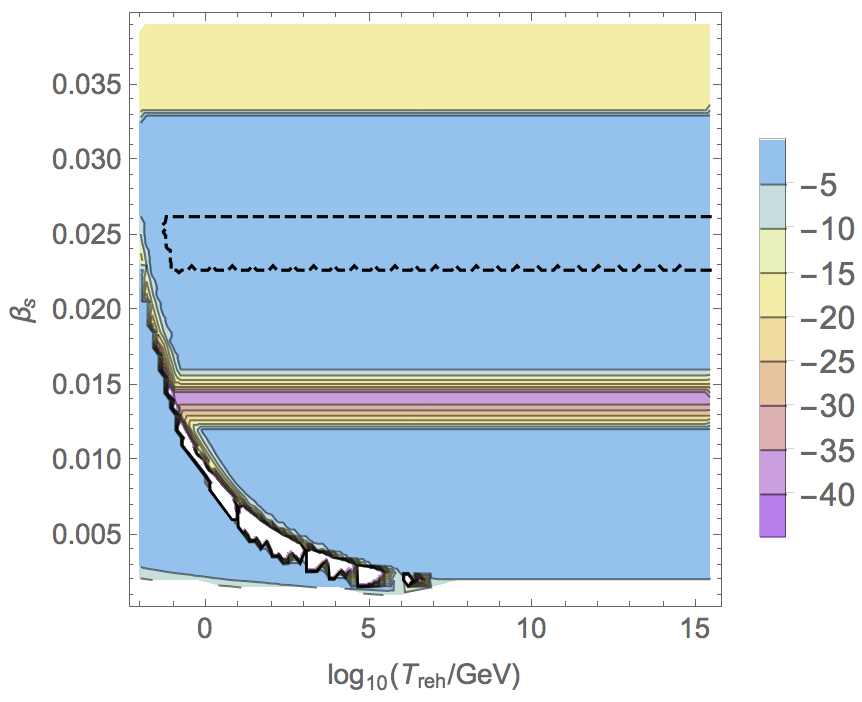}
\caption{Contour of the logarithm of the fraction of the PBH energy density in that of dark matter, $\log_{10}f_{\text{PBH}}$. Except fine-tuned points explained below, no region can explain the whole dark matter abundance.  Parameters are set as $\ns = 0.96, \alphas=0$, and $H_{\text{MD}}=10^{13}\text{GeV}$. The black dashed contour satisfies $f_{\text{PBH}}^{\text{LIGO}}=10^{-3}$ and the region inside it can explain the BH merger rate implied by LIGO/Virgo gravitational wave detection. Note that the integration region of $f_{\text{PBH}}^{\text{LIGO}}$ is restricted to the LIGO/Virgo range (see Eq.~\eqref{f_LIGO}), while $f_{\text{PBH}}$ does not have such a restriction (see Eq.~\eqref{f_integrate}).}
 \label{fig:f_contour}
  \end{figure}
\begin{figure}[tbh]
 \centering
   \subcaptionbox{\label{sfig:P(fLIGO)}}
{\includegraphics[width=0.48\columnwidth]{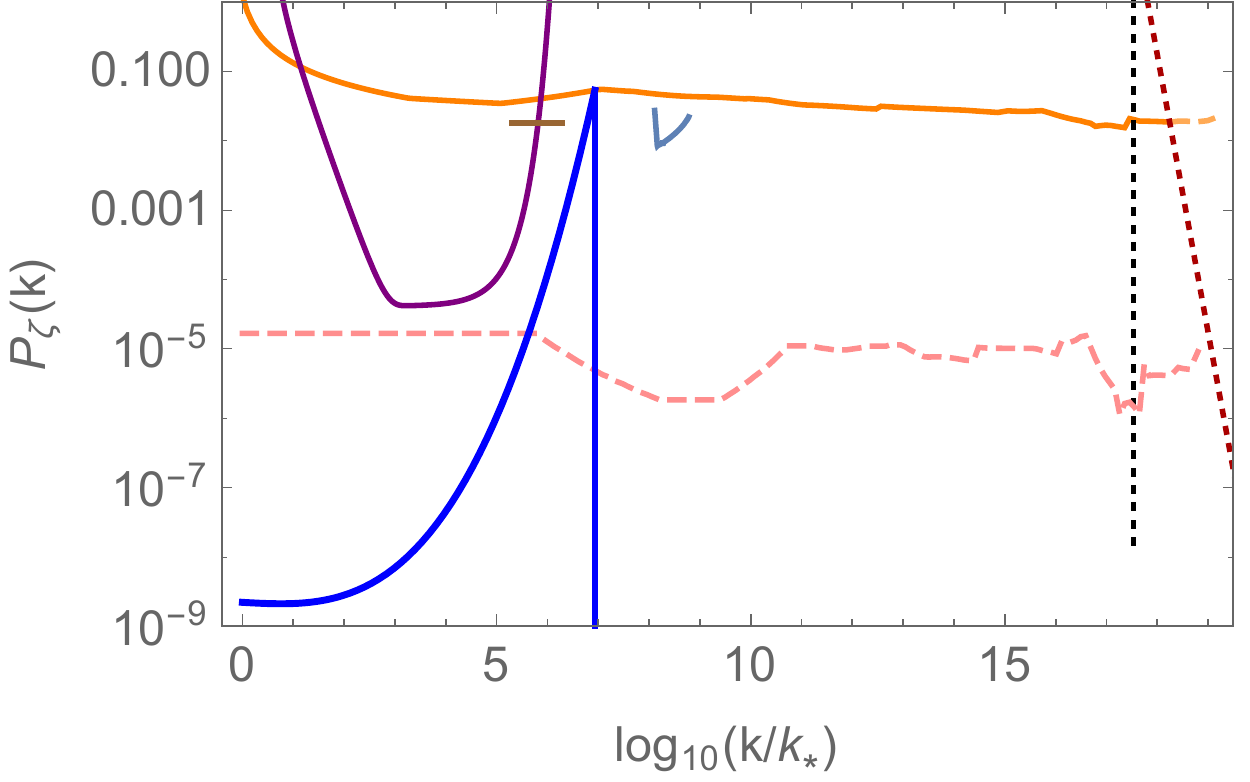}}~
  \subcaptionbox{\label{sfig:fLIGO}}
{\includegraphics[width=0.48\columnwidth]{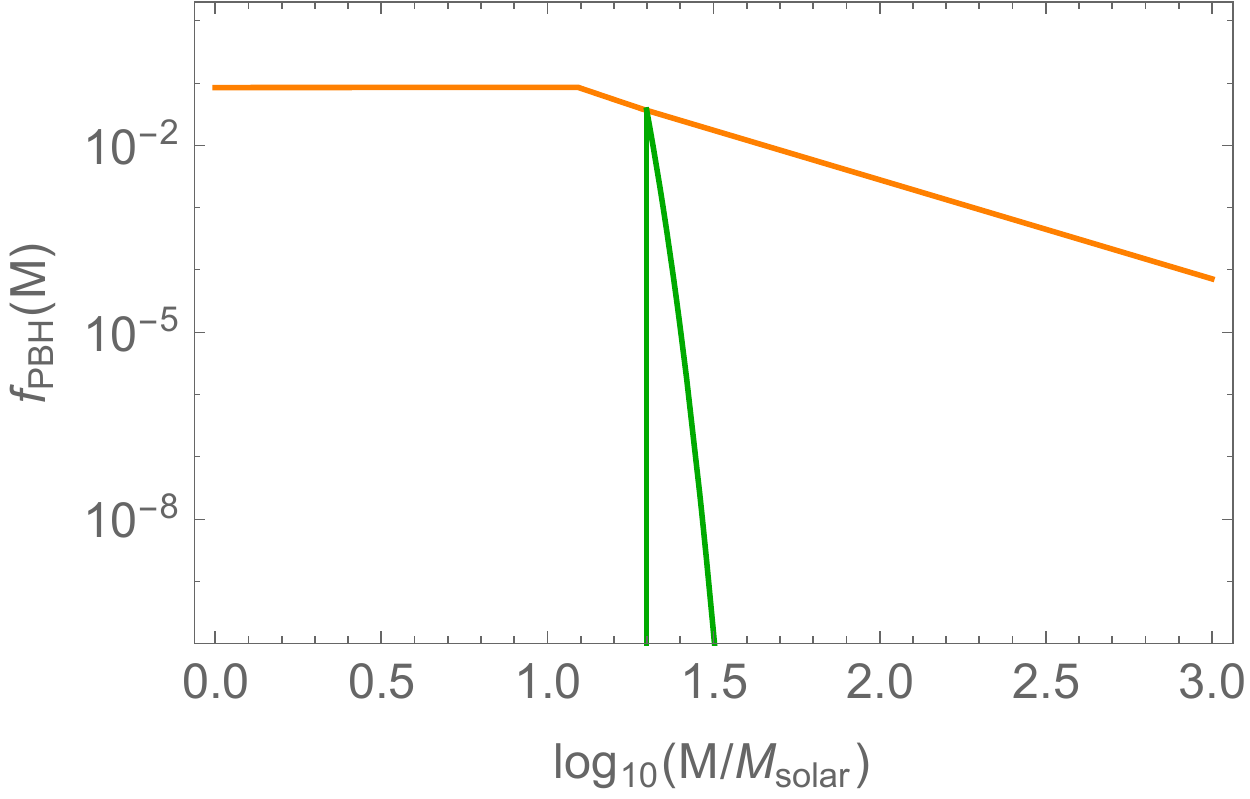}}~
  \caption{(\subref{sfig:P(fLIGO)}) An example of the power spectrum $P_\zeta (k)$ (blue solid line) which realizes $f_{\text{PBH}}^{\text{LIGO}}=1.2 \times 10^{-3}$. The parameters are $\ns=0.96, \alphas = 0, \betas = 0.026$.  The orange line is the constraint on $f_{\text{PBH}}(M)$ for RD. The constraint line for MD (light red dashed line) is not effective at these scales.  Other solid lines show the constraints of CMB, BBN, and PTA from left to right.  The black dotted line shows the reheating at $T_{\text{R}}=10^9 \text{GeV}$, and the dark red dotted line shows the minimum $k$ which becomes nonlinear during the MD era. (\subref{sfig:f=1}) The corresponding mass function $f_{\text{PBH}}(M)$ (green solid line).  The orange line is the constraint (for the monochromatic mass case). }
 \label{fig:fLIGO}
\end{figure}
\begin{figure}[tbh]
 \centering
\includegraphics[width=0.5 \columnwidth]{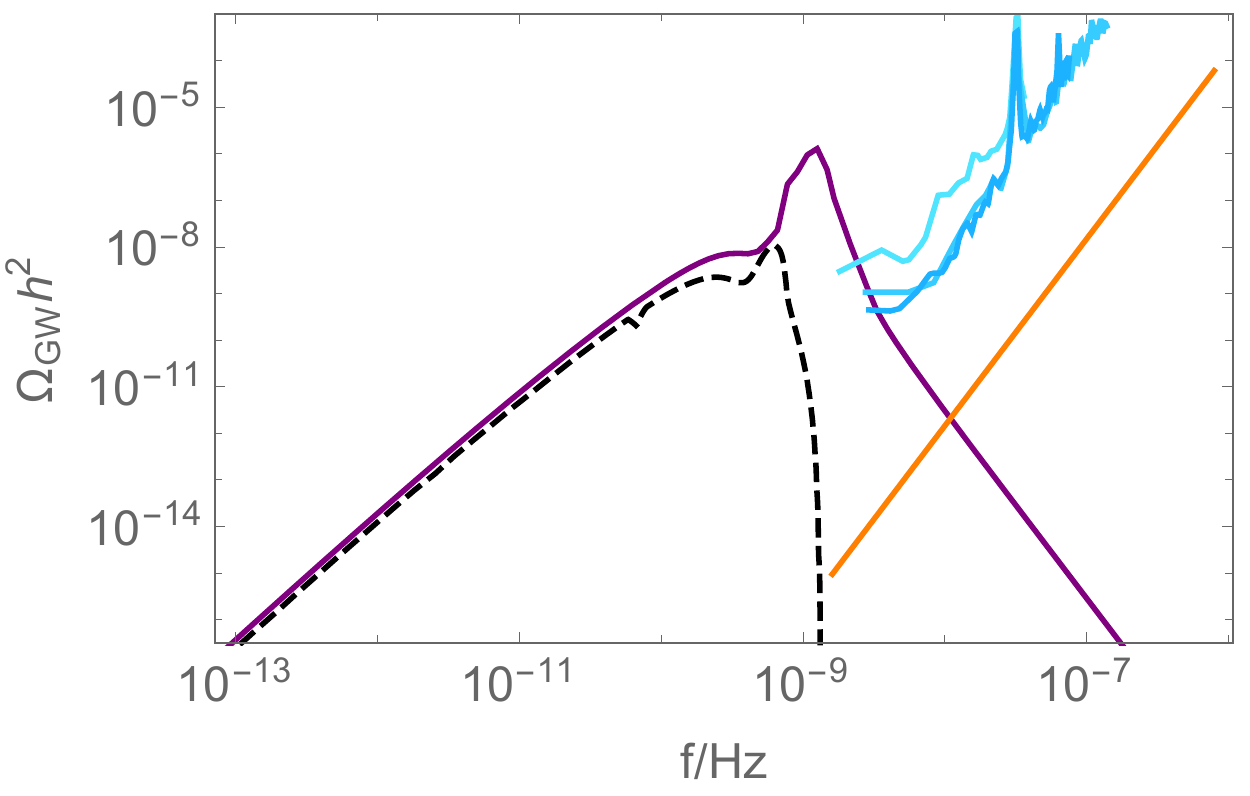}
\caption{The gravitational wave spectrum induced by the enhanced curvature perturbations which produce PBHs explaining the LIGO/Virgo BBH merger rate (corresponding to Fig.~\ref{fig:fLIGO}).  The black dashed  and purple solid lines are the cases of a sharp cutoff and a power-law cutoff with the spectral index $-2$, respectively.  The current PTA constraints (EPTA~\cite{Lentati:2015qwp}, NANOGrav~\cite{Arzoumanian:2018saf}, PPTA~\cite{Shannon:2015ect}) and the future one (SKA~\cite{5136190}) are also shown by the bluish lines and the orange line, respectively.}
 \label{fig:GW_LIGO}
  \end{figure}

In Fig.~\ref{fig:f_contour} we show the contour of
$\log_{10}f_{\text{PBH}}$ on the $(\betas, T_\text{R})$ plane.  The
contour of $\log_{10} f_{\text{PBH}}^{\text{LIGO}}=-3$ is also
superimposed on the figure as the black dashed line.  Other
parameters are set to the following values: $\ns = 0.96, \alphas=0$,
and $H_{\text{MD}}=10^{13}\text{GeV}$.  Changes of $\ns$ and $\alphas$
slightly shift the value of $\betas$ which gives the same
$f_{\text{PBH}}$, and a change of $H_{\text{MD}}$ controls the maximum
$T_{\text{R}}$ but hardly affects $f_{\text{PBH}}$.  That is why we
choose the $(\betas, T_{\text{R}})$ plane in Fig.~\ref{fig:f_contour}.
In the figure, the upper right region is dominated by the PBHs
produced in the RD era, while the lower left region is dominated by
those produced in the MD era. 
The white band corresponds to the scale of reheating where the
constraints for the RD and MD eras switch.  The band and the nearby
non-smooth structures should not be taken seriously because this is
related to the condition when we end the scan of $k$.  The pink
horizontal band near $\betas \simeq 0.014$ where $f_{\text{PBH}}$
becomes tiny is due to exclusions by the PTA constraint.  On the other
hand, the yellow band above $\betas \simeq 0.33$ is due to the BBN/CMB
constraint on the scalar perturbations.  In these regions, the
constraints are approximately independent of $T_{\text{R}}$ simply
because the PBH forms almost instantaneously in the RD era when the relevant mode enters the horizon, and information such as when the MD era ended is irrelevant.
On the other hand, PBH formation in the MD era depends on $T_{\text{R}}$ since it takes some finite time for the seed perturbations to become nonlinear.

Within the black dashed line, 
$10^{-3}\leq f_{\text{PBH}}^{\text{LIGO}}\leq 10^{-2}$ is satisfied,
which explains~\cite{Sasaki:2016jop} the BH merger rate expected from the LIGO/Virgo
event, $2$--$53 \, \text{Gpc}^{-3} \text{yr}^{-1}$~\cite{Abbott:2016nhf}.  Let us see more
closely an example which realizes
$10^{-3}\leq f_{\text{PBH}}^{\text{LIGO}}\leq 10^{-2}$.
Fig.~\ref{fig:fLIGO}\subref{sfig:P(fLIGO)} shows the power spectrum of the primordial curvature
perturbations $P_{\zeta}(k)$ which results in the mass function $f_{\text{PBH}}(M)$ realizing $f_{\text{PBH}}^{\text{LIGO}}=1.2 \times 10^{-3}$ shown in Fig.~\ref{fig:fLIGO}\subref{sfig:fLIGO}.
For the parameter space that explains the merger rate, we have $f_{\text{PBH}}^{\text{LIGO}} \simeq f_{\text{PBH}}$ because the integral is dominated at smaller scales.
This shows that in the PBH scenario for the merger rate, the dark matter abundance explained by PBHs is at most a percent level.

So far, we assumed a sharp cutoff of the curvature perturbations resulting from sudden end of inflation, which makes it easy to circumvent the PTA constraints. Let us relax this assumption and consider a milder cutoff \textit{e.g.}~by a power-law $k^n$. It is found in Ref.~\cite{Inomata:2016rbd} that the power $n$ must be smaller than about $-2$.
Although we have shown the PTA constraints on the curvature perturbation in Fig.~\ref{fig:fLIGO}\subref{sfig:P(fLIGO)} adopted from Ref.~\cite{Carr2018prep}, it is more appropriate to calculate the induced gravitational wave spectrum and compare it with PTA constraints directly because even the monochromatic curvature perturbations induce the secondary gravitational waves with a finite width.
This is done in Fig.~\ref{fig:GW_LIGO} for the sharp cutoff case (black dashed line) and for the power-law case (purple solid line) using analytic formulas of Ref.~\cite{Kohri:2018awv}.
We see that the power-law cutoff with the index $-2$ is marginally excluded by PTA experiments (bluish lines).
In a realistic model, the shape of the curvature spectrum $\mathcal{P}_\zeta (k)$ will be different from ours, in particular around the peak, so it may or may not be excluded by the PTA constraints.  Even so, such a spectrum will be unambiguously tested by SKA (orange line) unless the spectrum is cut off by the step function.

Next, we consider the possibility to explain dark matter by PBHs.  
Note that there is a small gap in the constraints on
$f_{\text{PBH}}(M)$ in the asteroid-mass range
$M\simeq 4\times10^{-17}M_\odot$.  We consider parameter sets such
that the corresponding mass function hits the gap where the constraint
is mild in a MD era. This requires tuning of $\betas$ for given
$\ns, \alphas$ and $T_{\text{R}}$. For a higher $T_{\text{R}}$, the
magnitude of $P_\zeta (k)$ required to produce a fixed
$f_{\text{PBH}}(M)$ is lower.  However, when $T_{\text{R}}$ is too
high, the asteroid mass scale is out of the MD era, and thus there is an
optimum $T_{\text{R}}$ around $T_{\text{R}}\simeq 10^4 \, \text{GeV}$
to obtain a large $f_{\text{PBH}}$.
  
However, it is hard to find a parameter set which realizes
$f_{\text{PBH}}=1$ and does not touch the constraint at all.  Remember
that the constraints displayed in Fig.~\ref{fig:f(M)constraints} are derived by assuming monochromatic
mass functions.  Therefore, we allow intersections of the mass function and
the constraints, and fix the minimum PBH mass to
$M=3.69\times 10^{-17}M_\odot$ (the high-mass boundary of the
extra-galactic gamma-ray background constraint) to enhance
$f_{\text{PBH}}$.  Instead, we adopt the prescription introduced in
Ref.~\cite{Carr:2016drx} to check whether it is really excluded.  To
this end, we have to calculate the integrals of $f_{\text{PBH}}(M)$ in
the regions where the constraint function is a monotonically
increasing or decreasing function. (These integrals to check
  exclusion should not be confused with the integral to obtain the
  total PBH fraction $f_{\text{PBH}}$ which we always do.)

\begin{figure}[tbh]
 \centering
\includegraphics[width=0.5 \columnwidth]{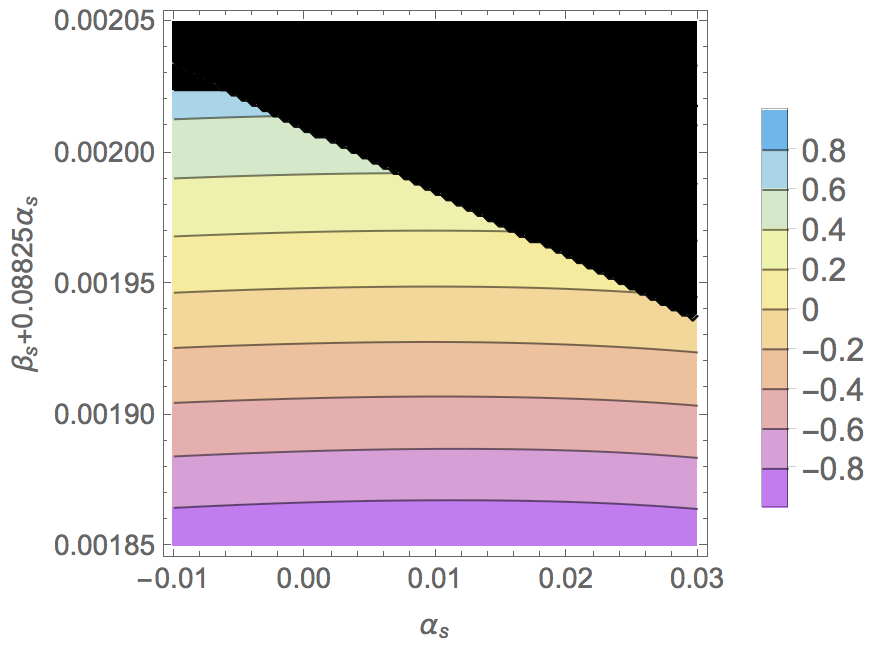}
  \caption{Contour of the logarithm of the fraction of the PBH energy density in that of dark matter, $\log_{10}f_{\text{PBH}}$, on the $(\alphas, \betas + 0.08825 \alphas)$ plane (for visibility; the plotted region is a thin strip in the ($\alphas, \betas$) plane).  Parameters are set as $\ns = 0.96, T_{\text{R}}=10^4 \text{GeV}$, and $H_{\text{MD}}=10^{13}\text{GeV}$.
  The black domains are excluded by either the femto-lensing of gamma-ray burst~\cite{Barnacka:2012bm} or caustic crossing~\cite{Oguri:2017ock} using the prescription for an extended mass function in Ref.~\cite{Carr:2016drx}. In the computation, the minimum mass is taken as $3.69 \times 10^{-17} M_\odot$. See the text for details.}
 \label{fig:f=1_contour}
\end{figure}
\begin{figure}[tbh]
 \centering
   \subcaptionbox{\label{sfig:P(f=1)}}
{\includegraphics[width=0.48\columnwidth]{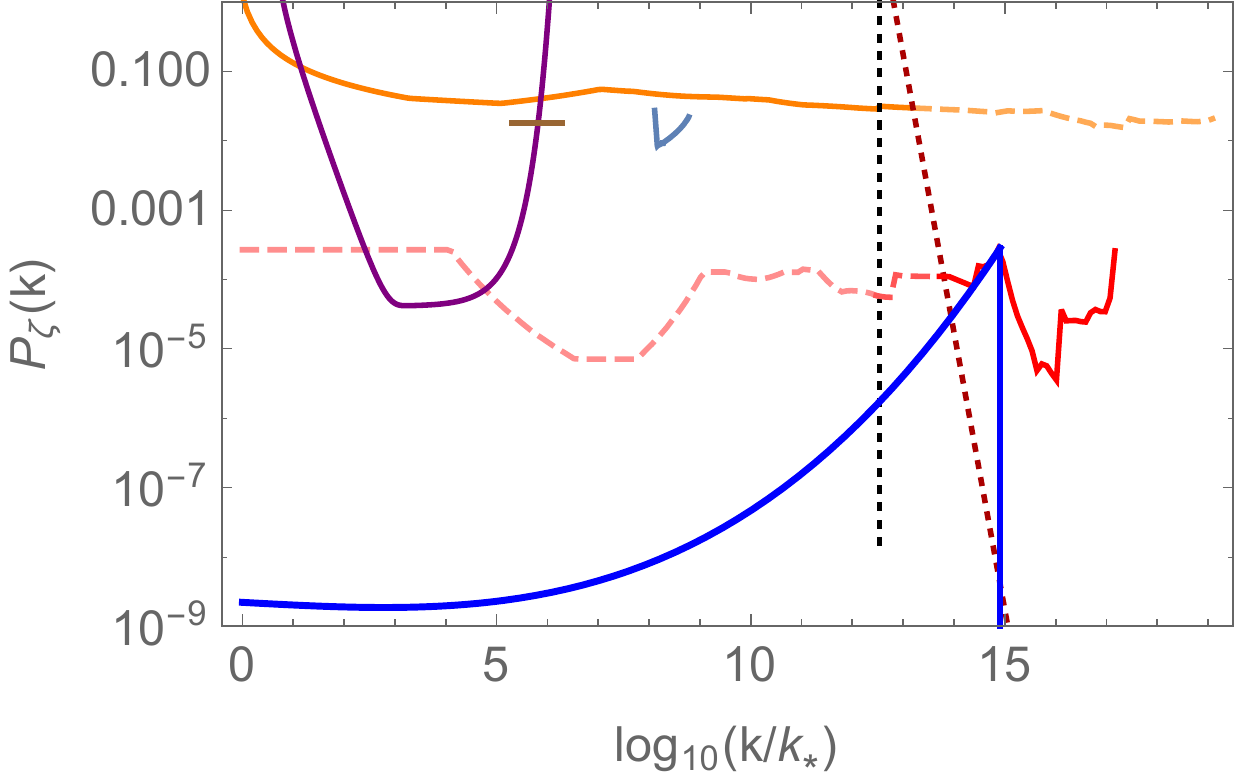}}~
  \subcaptionbox{\label{sfig:f=1}}
{\includegraphics[width=0.48\columnwidth]{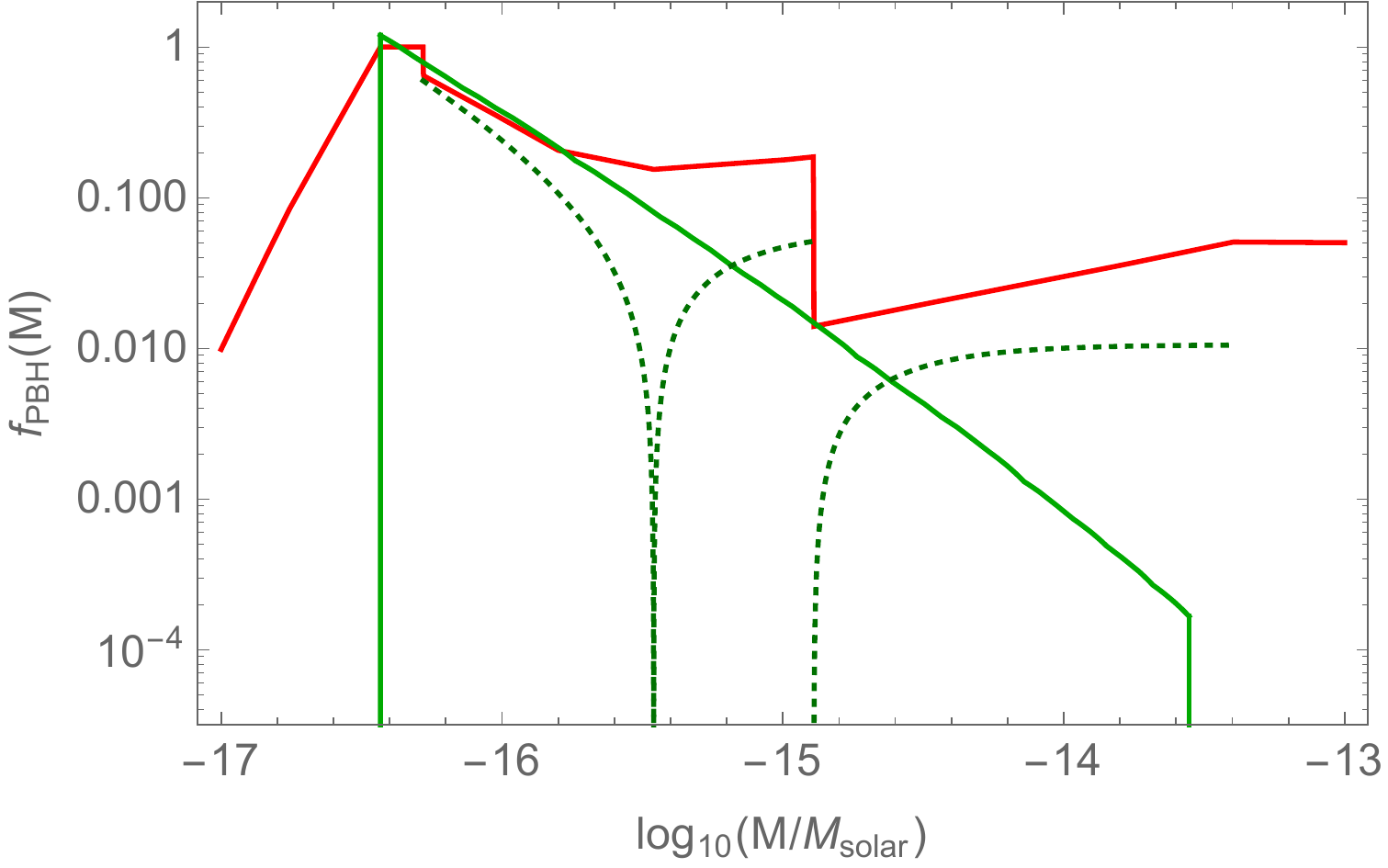}}~
  \caption{(\subref{sfig:P(f=1)}) An example of the power spectrum $P_\zeta (k)$ (blue solid line) which realizes $f_{\text{PBH}}=1.00$. The parameters are $\ns=0.96, \alphas = 0, \betas = 0.0019485$.  The orange and red lines are the constraints on $f_{\text{PBH}}(M)$ for RD and MD respectively. Other solid lines show the constraints of CMB, BBN, and PTA from left to right.  The black dotted line shows the reheating at $T_{\text{R}}=10^4 \text{GeV}$, and the dark red dotted line shows the minimum $k$ which becomes nonlinear during the MD era. (\subref{sfig:f=1}) The corresponding mass function $f_{\text{PBH}}(M)$ (green solid line).  The red line is the constraint for the monochromatic mass case.  The dotted dark-green lines are integrals of $f_{\text{PBH}}(M)$ in the region where the identical constraint is monotonically increasing or decreasing. The fact that the dotted dark-green lines are always below the red line shows that this mass function is not excluded (see the text and Ref.~\cite{Carr:2016drx} for details).}
 \label{fig:f=1}
\end{figure}
\begin{figure}[thb]
 \centering
\includegraphics[width=0.5 \columnwidth]{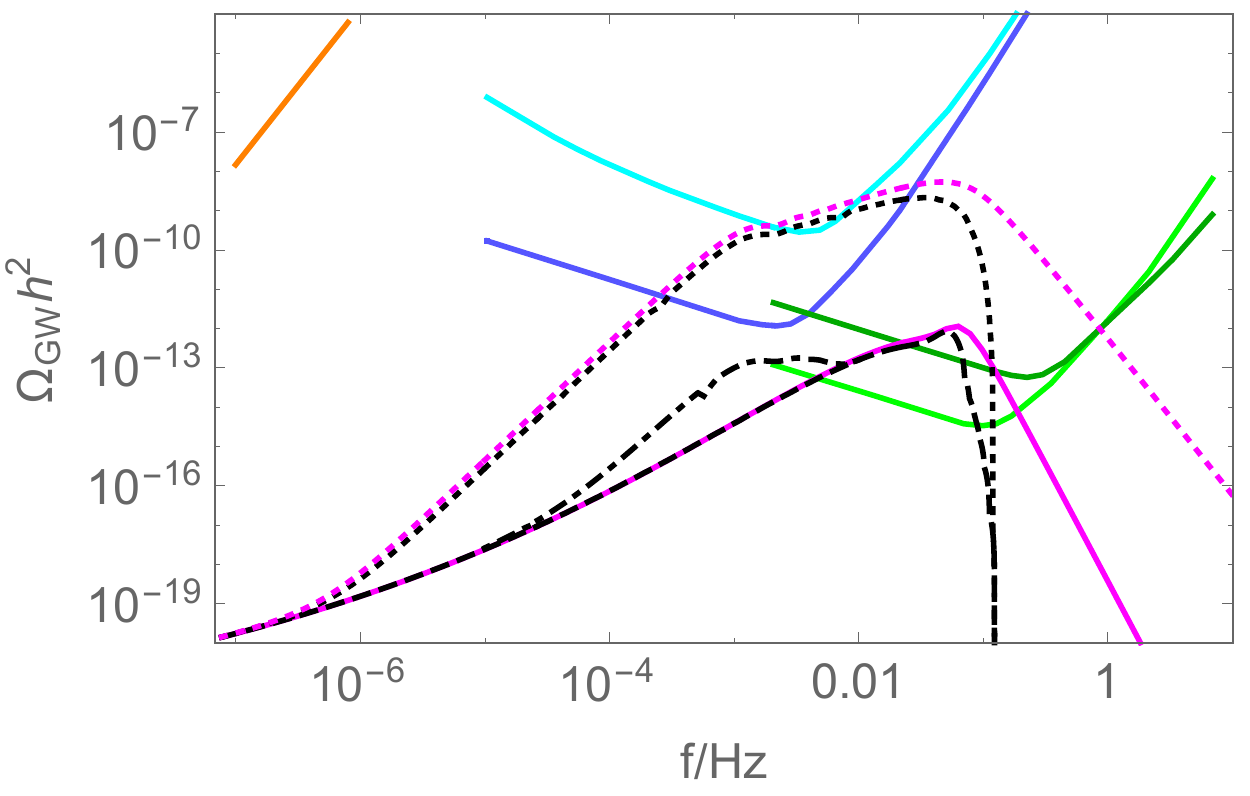}
\caption{The gravitational wave spectrum induced by the enhanced curvature perturbations which produce PBHs explaining the dark matter abundance (corresponding to Fig.~\ref{fig:f=1}).  
The black dashed line and magenta solid line represent the standard contribution present in the radiation-dominated era for the sharp cutoff and power-law cutoff cases respectively.
These are enhanced in the matter-dominated era to become the black dotted line and magenta dotted line respectively.
However, these enhancements are due to extrapolation of the linear formula into the non-linear regime.
The black dot-dashed line is conservative in the sense that all the contributions in the non-linear regime are neglected.
For comparison, the sensitivity curves of future observations, SKA~\cite{5136190}, eLISA~\cite{Seoane:2013qna}, LISA~\cite{2017arXiv170200786A},  BBO~\cite{Harry:2006fi}, and DECIGO~\cite{Seto:2001qf} are shown by the orange, cyan/blue, light green, and green lines, respectively.}
 \label{fig:GW_DM}
  \end{figure}

We find the following value 
 of $\betas$ results in
$f_{\text{PBH}}\simeq 1$,
\begin{align}
  \betas \simeq 0.00195 -0.08825 \alphas, \label{ab-relation}
\end{align}
for $\ns = 0.96$ and $T_{\text{R}}=10^4$ GeV.  The contour plot of
$f_{\text{PBH}}$ on the $(\alphas, \betas + 0.08825 \alphas)$ plane
(for visibility; the plotted region is a thin strip on the
($\alphas, \betas$) plane.)  is given in Fig.~\ref{fig:f=1_contour}.
The black domain is excluded either by the femto-lensing of gamma-ray
burst~\cite{Barnacka:2012bm} or by caustic
crossing~\cite{Oguri:2017ock}.  The power spectrum and the mass
function of an example ($\alphas=0$ and $\betas=0.0019485$) realizing
100\% dark matter are shown in Fig.~\ref{fig:f=1}.
Fig.~\ref{fig:f=1}\subref{sfig:P(f=1)} shows $P_\zeta (k)$ and
Fig.~\ref{fig:f=1}\subref{sfig:f=1} shows $f_{\text{PBH}}(M)$.

Again, the sharp cutoff assumption for the curvature perturbations is conservative for explaining the dark matter abundance, but aggressive for circumventing the constraints.
We can relax the assumption and consider \textit{e.g.} the power-law cutoff proportional to $k^n$.  Since the slope of the constraint on $\mathcal{P}_\zeta (k)$ from the extragalactic gamma-ray background is roughly $-3$, the power-law cutoff must satisfy $n \lesssim -3$.
This may be hard to be realized in a concrete setup.  However, in the presence of such a power-law tail, the dark matter abundance can be explained by curvature perturbations with a smaller peak, which then allows a larger power-law index. 
For completeness, we also show the spectra of the gravitational waves induced by the enhanced curvature perturbations in Fig.~\ref{fig:GW_DM}.
We see that BBO (light green line) and DECIGO (green line) can detect the induced gravitational waves even if the spectrum is sharply cut off (black dashed line).
Moreover, eLISA (cyan line) and LISA (blue line) may be able to detect them if the gravitational waves are sufficiently enhanced in the MD era (dotted lines).

\section{Summary and conclusions}\label{sec:conclusion}
In this paper, we have considered the power spectrum of the curvature
perturbation with running spectral indices parametrized by the
 parameters $\ns$, $\alphas$, and $\betas$ to produce PBHs,
varying also the reheating temperature $T_{\text{R}}$ which
parametrizes the onset of the radiation-dominated Universe after the
early matter-dominated Universe.  We have shown that it is possible to
explain the abundance of 100$\%$ DM or to fit the
merger rate of the BBH observed from the LIGO/Virgo
gravitational wave detections, utilizing the PBHs.  It is notable that those fittings are
possible even if we consider the serious suppression of the production
rates of the PBHs due to the conservation of angular momentum for
non-relativistic particles in the matter dominated
Universe~\cite{Harada:2017fjm}. For the BH merger rate, we need
$10^{-3} \lesssim f_{\text{PBH}}^{\text{LIGO}} \lesssim 10^{-2}$ which
is obtained if $0.023 \lesssim \betas \lesssim 0.026$ is realized (see
Fig.~\ref{fig:f_contour}) (a precise value depends on $\ns$ and
$\alphas$). Such a value of $\betas$ can be probed by future
observations of the cosmological 21cm line emissions and the
polarization of CMB~\cite{Kohri:2013mxa}.   For the 100\% DM abundance,
which is realized for the asteroid-mass ($\sim 10^{-17} M_\odot$) of
PBHs, $\alphas$ and $\betas$ have to be precisely correlated with each
other for fixed $\ns$ as shown in Eq.~\eqref{ab-relation} and
Fig.~\ref{fig:f=1_contour} with $T_{\text{R}}=10^4\,\text{GeV}$.
 
A lot of aspects of our work can be extended in future as our
limitations are discussed in Sec.~\ref{sec:running}.  It will be worth
building concrete inflation models which realize characteristic
features, including the running spectral indices, of the power
spectrum studied in this paper.  This enables us to discuss
 \textit{e.g.}~how quickly the running inflation ends, whether ultra
slow-roll features appear~\cite{Motohashi:2017kbs}, and how the
Gaussian assumption is reasonable~\cite{Franciolini:2018vbk}, etc.  Since the power spectrum
with running parameters is relatively simple, these concrete models
may hopefully be simple as well.

\section*{Acknowledgments}
This work is supported in part by JSPS Research Fellowship for Young
Scientists (TT) and JSPS KAKENHI grant Nos.~JP17J00731 (TT),
JP17H01131 (KK), 26247042 (KK), and MEXT KAKENHI Grant Nos.~JP15H05889
(KK), and JP16H0877 (KK).

\small

\bibliographystyle{utphys}
\bibliography{pbh}

\end{document}